\documentclass[12pt]{article}
\usepackage{amsfonts, amsmath, amssymb, amsfonts, amscd, bm, booktabs, xcolor, enumitem, epstopdf, mathabx, hyperref, graphicx}
\usepackage{dcolumn}
\newcolumntype{.}{D{.}{.}{-1}}
\newcolumntype{d}[1]{D{.}{.}{#1}}
\usepackage[top=.8in, left=.5in, right=.5in,
bottom=.8in]{geometry}
\RequirePackage{natbib}
\usepackage[compact]{titlesec}
\usepackage{booktabs}
\usepackage{url} 

\usepackage{color}

\newtheorem{assumption}{Assumption}
\newtheorem{remark}{Remark}

\newtheorem{theorem}{Theorem}
\newtheorem{corollary}{Corollary}

\usepackage{caption}
\usepackage{subcaption}
\usepackage[linesnumbered, ruled, vlined]{algorithm2e}

\expandafter\def\expandafter\normalsize\expandafter{\normalsize\setlength\abovedisplayskip{0pt}}
\expandafter\def\expandafter\normalsize\expandafter{\normalsize\setlength\belowdisplayskip{0pt}}
\expandafter\def\expandafter\normalsize\expandafter{\normalsize\setlength\abovedisplayshortskip{0pt}}
\expandafter\def\expandafter\normalsize\expandafter{\normalsize\setlength\abovedisplayshortskip{0pt}}

\begin{document}
\pagestyle{plain}

\def\spacingset#1{\renewcommand{\baselinestretch}%
{#1}\small\normalsize} \spacingset{1.5}


  \title{Efficient surrogate-assisted inference for patient-reported outcome measures with complex missing mechanism}
  \author{Jaeyoung Park \thanks{
  	Booth School of Business, University of Chicago}\\
  	\and
  	Muxuan Liang\thanks{
  	Department of Biostatistics, University of Florida}\\
  	\and
  	Ying-Qi Zhao\thanks{
  	Public Health Sciences Divisions, Fred Hutchinson Cancer Center}\\
  	\and
  	Xiang Zhong\thanks{
  	Department of Industrial and Systems Engineering, University of Florida}}
  \date{}
  \maketitle

\thispagestyle{empty}

\abstract{
Patient-reported outcome (PRO) measures are increasingly collected as a means of measuring healthcare quality and value. The capability to predict such measures enables patient-provider shared decision making and the delivery of patient-centered care. However, PRO measures often suffer from high missing rates, and the missingness may depend on many patient factors. Under such a complex missing mechanism, developing a predictive model for PRO measures with valid inference procedures is challenging, especially when flexible imputation models such as machine learning or nonparametric methods are used. Specifically, the slow convergence rate of the flexible imputation model may lead to non-negligible bias, and the traditional missing propensity, capable of removing such a bias, is hard to estimate due to the complex missing mechanism. To efficiently infer the parameters of interest, we propose to use an informative surrogate that enables a flexible imputation model lying in a low-dimensional subspace. To remove the bias due to the flexible imputation model, we identify a class of weighting functions as alternatives to the traditional propensity score and estimate the low-dimensional one within the identified function class. Based on the estimated low-dimensional weighting function, we construct a one-step debiased estimator without using any information of the true missing propensity. We establish the asymptotic normality of the one-step debiased estimator. Simulation and an application to real-world data demonstrate the superiority of the proposed method.
}

\newcommand{\n}{\noindent}
\maketitle

\section{Introduction}
\label{sec:intro}

Patient-reported outcome (PRO) measures are increasingly collected before and after an intervention or a treatment as a means of measuring healthcare quality and value, which is an important step toward patient-centered care. Knowing the measure goes up or down alone might not be sufficient to determine the effectiveness of the intervention. More importantly, whether the measure has changed with a sufficiently large margin, known as the minimally clinically important difference (MCID), needs to be evaluated. If the intervention is an elective surgery, identifying patients at risk of not achieving an MCID, particularly before the surgery, is important for pre-surgical decisions. There is a growing interest in applying machine learning techniques to predict whether a patient is likely to achieve an MCID before their surgery and identify predictive factors associated with post-surgical PRO measures. 

The increasing adoption of electronic health record (EHR) systems has provided unprecedented opportunities to learn an interpretable model for predicting PRO measures using massive observational data. Although the volume of observational data is large, the quality of such observational data may be uncertain. One of the major difficulties  is missing data, especially missing the outcome data. In our motivating example, the MCIDs can only be observed from the participants who take both pre- and post-surgical surveys. The participants who completed both surveys may only account for a small portion (e.g., 1/3) of the participants whose EHR data is available, according to the response rate reported in literature  \citep{ho2019improving, pronk2019response} and from our own data. Unfortunately, low survey response rates are not uncommon in healthcare and other service industries. In this work, our objective is to develop an interpretable predictive model for the outcome subject to missing. Specifically, we aim at developing a linear prediction model by minimizing the deviance of a generalized linear model (GLM) with a valid inference procedure for the coefficients under possible model misspecification.

Many approaches have been developed to deal with missing outcomes under the assumption of missing at random (MAR) \citep{kang2007demystifying}. One seminal work is the propensity inverse weighting approach \citep{rosenbaum1983central, horvitz1952generalization}. For this approach, one first estimates the probability of missing w.r.t the covariate (also called the propensity) and then uses the inverse of the estimated propensity to adjust for the selection bias. When the propensity is poorly estimated, the propensity inverse weighting methods may not perform well. Another major type of approach is known as  imputation. This approach first learns an imputation model using the fully observed part of the data; then, imputes the missing outcomes with the predicted values; and finally, refits the predictive model based on the imputed outcomes \citep{rubin2004multiple}. When the estimated imputation model is misspecified, the refitted predictive model may also be biased. To maintain robustness against the possible misspecification in the propensity and the imputation models, one possible solution is to use the doubly robust methods \citep{robins1994estimation}. The doubly robust methods that incorporate both the propensity score and the imputation models can lead to a consistent estimate for the outcome as long as either  model is correctly specified \citep{tan2006distributional, tan2010bounded, qin2008efficient, qin2007empirical, rubin2008empirical, cao2009improving, han2012note, rotnitzky2012improved, han2016doubly}.

Statistical inference for the parameters in predictive modeling with outcome missingness is also challenging. In particular, when the missing mechanism is dependent on multiple covariates through a nonlinear relationship, an unbiased estimator for the missing propensity with a fast convergence rate may be infeasible. For the inverse weighting approaches and the doubly robust methods, a parametric model for the propensity may not capture the potential non-linearity. To ensure an unbiased propensity estimate, nonparametric regressions and machine learning methods have been adopted. These methods may lead to a slower convergence rate and hinder the inference of the parameters in the predictive model, especially when the number of the covariates is large. When the number of the covariates is small, to address the slow convergence rate, the double machine learning approach was proposed in \cite{victor2018}. They adopted a cross-fitting algorithm using a doubly robust formulation and proposed to estimate both the propensity and the imputation model using nonparametric or machine learning methods. They proved that, as long as the product of the convergence rates of the propensity and imputation estimates is smaller than $n^{-1/2}$,  a valid inference procedure for the parameters in the predictive model is possible, where $n$ represents the sample size. However, the large number of the covariates and not meeting the smoothness condition on the true propensity may negate the required rate condition.

To help address the above statistical inference challenge due to the presence of a large number of  covariates, one possible strategy is to leverage a surrogate outcome. The surrogate outcomes herein are defined as alternative clinical outcomes that are likely to predict the clinical benefit of primary interest. In our motivating example,  the MCID of the global physical health T-score in the Patient-Reported Outcomes Measurement Information System (PROMIS) survey is a well acknowledged measurement for evaluating surgery benefit. There are other PRO measures collected that represent different but related mental or physical health performances that can be considered as surrogate outcomes. In many applications, a surrogate outcome can help improve the efficiency or overcome the difficulties due to complex missing mechanisms. In the application of causal inference \citep{prentice1989surrogate, frangakis2002principal, fleming1994surrogate, cheng2018efficient, anderer2021adaptive}, a surrogate can be used to improve the efficiency of estimating the average treatment effect (ATE). In the application of semi-supervised inference, under the assumption of missing completely at random (MCAR), \cite{hou2021surrogate} showed that a surrogate can help infer the predicted risk derived from a high-dimensional working model even when the true risk prediction model depends on multiple covariates. However, their approach cannot be applied under the assumption of missing at random (MAR), which is the setting we need to deal with.

In this work, we focus on how to use  surrogate outcomes to develop interpretable predictive models with outcome missingness. The parameter of interest herein is defined as the minimizer of the deviance under a GLM with  possible model misspecification. We propose a concept of an informative surrogate, defined as a surrogate outcome that enables a low-dimensional imputation model conditional on the surrogate and the covariates (i.e., the imputation model lies in a low-dimensional subspace generated by the surrogate and covariates). Under the MAR assumption, we exploit the role of this informative surrogate to 1) allow  for a low-dimensional imputation model under a large number of covariates; 2) avoid estimating the complex missing propensity. To harvest the potential benefit brought by informative surrogate outcomes, we propose the following procedure. First, we estimate a flexible imputation model (e.g., using kernel regression or basis expansion) in a reduced subspace that is constructed by leveraging the information from informative surrogate outcomes. Subsequently, we can impute the missing outcomes and obtain an initial estimator for the parameters of interest. Then, we bypass the estimation of the complex missing propensity and instead estimate a low-dimensional weighting function based on the reduced subspace to adjust for the possible bias due to the estimated imputation model. Finally, a one-step debiased estimator for the parameters in the predictive model can be constructed. Both the point and interval estimates of the parameters can be obtained from the proposed procedure. We show that the proposed method can provide a valid inference procedure for the parameters of interest without requiring a consistent propensity estimation. In addition, when the true propensity lies in the same subspace as the imputation model, the proposed method leads to a semiparametric efficient estimator for the parameters in the predictive model. Extensive simulation and an analysis of real-world data are provided to demonstrate the superior performance of the proposed method.

The remainder of the paper is organized as follows. In Section~\ref{sec:method}, we define the parameter of interest and introduce our proposed method. In Section~\ref{sec:theory}, we demonstrate the theoretical validity of the proposed method. In Section~\ref{sec:sim}, we provide numerical studies to bolster the superiority of the proposed methods over other existing methods and methods without information of the surrogate. In Section~\ref{sec:real}, we apply the proposed method to derive a predictive rule to infer post-surgery improvement for joint replacement surgery patients. In Section~\ref{sec:disc}, we discuss possible future works.

\section{Method}
\label{sec:method}

 Let $\bm X$ be a $p$-dimensional covariate and $Y$ be a binary, categorical, or continuous outcome of interest. Without loss of generality, we choose a GLM as a working model for $E\left[Y\mid \bm X\right]$. Following the notation of exponential family distributions \citep{shao2003mathmetical}, a GLM assumes that $E\left[Y\mid \bm X\right]=b'(\bm X^\top\bm \beta)$, where $b'(\cdot)$, the derivative of function $b(\cdot)$, is a known link function. The parameter of interest, $\bm \beta$ is often defined as the minimizer of the deviance (or equivalently, the negative log-likelihood) under the working model, i.e.,
$
    \bm \beta^*=\arg\min E\left[\ell(\bm \beta)\right],
$
where
$
	\ell(\bm \beta)=b(\bm X^\top\bm \beta)-Y\bm X^\top\bm \beta.
$
If the working model is misspecified, i.e., $E\left[Y\mid \bm X\right]\not=b'(\bm X^\top\bm \beta^*)$, $\bm \beta^*$ that minimizes the deviance, a goodness-of-fit statistic, is still meaningful. For a linear working model, the link function $b'(t)$ is the identity function, and the function $b(t)=t^2/2$; the objective is equivalent to the least square. Notice that the parameter of interest $\bm\beta^*$ is defined under the full distribution where $Y$ and $\bm X$ are always observed. To ensure that $\bm\beta^*$ can be identified under the full distribution, we assume that
$b^{''}(\cdot)$ is always positive and $E[\bm X\bm X^\top]$ is positive definite.

For actual data, the outcome $Y$ can be missing. We collect the covariate $\bm X$, the outcome $Y$, the informative surrogate outcome $Z$, and the missing indicator $R$ from all samples. The missing indicator $R$ indicates whether $Y$ is observed ($R=1$) or not ($R=0$). 
We also assume that the surrogate $Z$ can be fully observed. Collectively, the observed data can be denoted as $(\bm X, Z, R, RY)$. To ensure the identifiability of $\bm\beta^*$ using the actual data, we assume that $Y\perp R\mid X, Z$.
 
 \subsection{First step: dimension reduction through informative surrogate}
\label{subsec:imputation}

In this section, we propose a two-step procedure under the assumption of $Y\perp R\mid \bm X, Z$. To start with, we formally define the concept of informative surrogate outcomes and introduce the required assumption for the identifiability of $\bm\beta^*$. 

An surrogate outcome $Z$ is informative if there exists a $(p+1)\times d$ matrix, $\bm \Gamma$, with orthogonal columns satisfying $Y\perp \widetilde{\bm X}\mid \bm \Gamma^\top \widetilde{\bm X}$ and $d<p$, where $\widetilde{\bm X}^\top=(Z, \bm X^\top)$. This definition implies that, conditioning on the surrogate outcome, the dimension of the space constructed by the covariates and the surrogate can be reduced to $d$, which is expected to be much smaller than $p$. The columns of $\bm \Gamma$ represent the reduced subspace. Thus, if the surrogate is informative, $Q(Z, \bm X):=E\left[Y\mid Z, \bm X\right]$ is a function lying in a low-dimensional subspace, i.e., there exists an unknown link function $g$ such that $Q(Z, \bm X)=g(\bm \Gamma^\top \widetilde{\bm X})$. Consequently, an efficient estimator to this low-dimensional imputation model may have a faster convergence rate than directing using the kernel regression to estimate $E\left[Y\mid \bm X\right]$ which is $p$-dimensional. 

Various existing methods can be employed to estimate the reduced subspace when the actual data is fully observable. The assumption $Y\perp \widetilde{\bm X}\mid \bm \Gamma^\top \widetilde{\bm X}$ is closely related to the (sufficient) dimension reduction literature. In the literature \citep{Li1991,cook2007, Xia2002, Xia2007, Ma2012, Ma2013}, the smallest space generated by the columns of $\bm \Gamma$ that satisfies $Y\perp \widetilde{\bm X}\mid \bm \Gamma^\top \widetilde{\bm X}$ is referred to as the central subspace. When the data is fully observed, the dimension reduction methods, such as the minimum average variance estimation (MAVE) \citep{Xia2002}, sliced-inverse regression (SIR) \citep{Li1991}, or semiparametric approaches in \cite{Ma2012, Ma2013}, can be directly applied to estimate the central subspace $\bm \Gamma$, and the estimated subspace is asymptotically normal. When there are multiple surrogate outcomes in the observed data, the above-mentioned methods (e.g., the MAVE) can be used to select candidate informative surrogate outcomes. For example, we can select the surrogate outcome that leads to the lowest reduced dimension.   

For our setting with the incomplete outcome data, to ensure the identifiability of $\bm\beta^*$ and $\Gamma$, traditional positivity assumption requires that $P(R=1\mid \widetilde{\bm X})>0$. In this work, instead of assuming the traditional positivity assumption, we assume a relaxed positivity assumption, $P(R=1\mid \bm \Gamma^\top \widetilde{\bm X})>0$. Under this assumption, if $Y\perp \widetilde{\bm X}\mid \bm \Gamma^\top \widetilde{\bm X}$, we can show that $Y\perp R\mid \bm \Gamma^\top \widetilde{\bm X}$, and 
$
	Q(Z, \bm X)=E\left[Y\mid Z, \bm X\right]=E\left[Y\mid \bm \Gamma^\top \widetilde{\bm X}\right]=E\left[Y\mid \bm \Gamma^\top \widetilde{\bm X}, R=1\right].
$
This implies that the conditional mean of $Y$ restricted to $R=1$ shares the same subspace with the unrestricted conditional mean. Thus, to estimate $\bm \Gamma$, we only need to apply these dimension reduction methods to the fully observed part of the data. Since $\bm\Gamma$ (and $Q(Z, \bm X)$) is identifiable, $\bm\beta^*$ is also identifiable under the relaxed positivity assumption.

After we obtain $\widehat{\bm \Gamma}$, we can use nonparametric regressions or machine learning methods to fit $Y$ w.r.t $\widehat{\bm \Gamma}^\top \widetilde{\bm X}$ to derive the unknown link function $g$ and estimate $\widehat{Q}(Z, \bm X)=\widehat{g}(\widehat{\bm \Gamma}^\top \widetilde{\bm X})$. Then, we obtain an initial estimator for $\bm \beta^*$ by minimizing
$
    \widehat{E}_n\left[b(\bm X^\top\bm \beta)-\widehat{g}(\widehat{\bm \Gamma}^\top \widetilde{\bm X})\bm X^\top\bm\beta\right],
$
or equivalently, solving the estimating equation
$$
    \widehat{E}_n\left[\left\{b'(\bm X^\top\bm \beta)-\widehat{g}(\widehat{\bm \Gamma}^\top \widetilde{\bm X})\right\}\bm X\right].
$$

Denote the solution as $\widehat{\bm \beta}$. Due to the slow convergence rate of nonparametric regressions or machine learning methods, the convergence rate of $\widehat{Q}$ is dominated by that of $\widehat{g}$. Subsequently,  $\widehat{\bm \beta}$ may suffer from the slow convergence rate of $\widehat{g}$. Thus, to obtain an estimator with a faster convergence rate, we need to remove the bias due to the estimation error of $\widehat{g}$, which will be discussed in the following section.

\begin{remark}
    The term $\widehat{g}(\widehat{\bm\Gamma}^\top \widetilde{\bm X})$ is expected to form a good prediction for $Y$. As an imputation model, it should be predictive of $Y$; however, there are at least two reasons that it may not be satisfactory. First, $\widehat{g}(\widehat{\bm\Gamma}^\top \widetilde{\bm X})$ depends on the surrogate outcome $Z$, which may not be available at the time of making the prediction and thus is not appropriate be treated as a covariate. Second, when the reduced dimension, i.e., the dimension of $\widehat{\bm\Gamma}^\top \widetilde{\bm X}$, is greater than $2$, it could be hard to interpret the model. 
\end{remark}

\subsection{Second step: debias using a low-dimensional weighting function}
\label{subsec:weighting}

In this section, we introduce the second step of the proposed method. Specifically, we propose to remove the bias due to the estimation error of $\widehat{g}$ using a low-dimensional weighting function. For ease of exposition, we focus on how to construct an improved estimator for $\beta_1^*$, which is the first coefficient in $\bm \beta^*$; the proposed method can be extended to infer $\bm u^\top\bm \beta^*$ for any $\bm u$. Thus, to get an improved estimator for $\bm\beta^*$, we can implement the proposed method for each coordinate of $\bm\beta^*$ and then ensemble these estimates to construct an estimator for $\bm\beta^*$.

To start with, we consider a class of estimating equations for $\beta_1^*$. We first derive the efficient influence function of $\beta_1^*$ without assuming any relationship between $Y$ and $Z$ given covariates $\bm X$. The details regarding the efficient influence function can be found in the Online Supporting Information. Motivated by the efficient influence function, we then consider the following class of estimating equations for $\beta_1^*$,
\begin{eqnarray}\label{eq:estimating_equation}
		\left\{S(\bm \beta; \pi, {Q})\right\}^\top \bm v,
\end{eqnarray}
where
\begin{eqnarray*}
    S(\bm \beta; \pi, {Q})&=&\left[\left\{b^\prime (\bm X^\top\bm \beta) -{Q}(Z,\bm X)\right\}+{\pi^{-1}(\bm X, Z)}R\left\{{Q}(Z,\bm X)-Y\right\}\right]\bm X.
\end{eqnarray*}
In this class of estimating equations, the first term $\left\{b^\prime (\bm X^\top\bm \beta)-{Q}(Z,\bm X)\right\}\bm X^\top \bm v$ is the estimating equation using ${Q}(Z,\bm X)$ as the imputation for all the outcomes. In the first step, we have obtained an imputation model $\widehat{Q}(Z,\bm X)$, which can be plugged into estimating equations~\eqref{eq:estimating_equation}. The second term $$\pi^{-1}(\bm X, Z)R\left\{Q(Z,\bm X)-Y\right\}\bm X^\top \bm v$$ can be interpreted as an efficiency augmentation term using a weighting function $\pi^{-1}(\bm X, Z)$. To construct an estimating equation for $\beta_1^*$, we can choose a specific $\bm v$ and $\pi^{-1}(\bm X, Z)$ in $\left\{S(\bm \beta; \pi, \widehat{Q})\right\}^\top \bm v$ with $\bm\beta$ replaced by $(\beta_1,\widehat{\bm\beta}_{-1})$, where $\widehat{\bm\beta}_{-1}$ is the sub-vector of $\widehat{\bm\beta}$, excluding the first coordinate. 

However, directly solving this estimating equation for an arbitrary choice of $\bm v$ and the weighting function may not lead to an improved estimator due to the estimation error of $\widehat{g}$. This estimation error affects the estimating equation via two paths. First, the estimating equation depends on $\widehat{Q}$, which is affected by the estimation error of $\widehat{g}$; second, the estimating equation depends on $\widehat{\bm\beta}_{-1}$, which is also affected by the estimation error of $\widehat{g}$.

In order to remove the estimation error of $\widehat{g}$, we need to remove the estimation errors of $\widehat{Q}$ and $\widehat{\bm\beta}_{-1}$. To remove the estimation error of $\widehat{\bm\beta}_{-1}$, we adopt the idea of de-correlated score \citep{ning2017}. The de-correlated score projects the score in a chosen direction such that the projected estimating equation is not affected by the estimation error of $\widehat{\bm\beta}_{-1}$. Following this idea, we choose the following $\bm v$ to achieve this goal. Let $\bm w^*$ be the minimizer of 
$$
	E\left[b^{''}(\bm X^\top\bm \beta^*)(X_1- \bm X_{-1}^\top \bm w)^2\right],
$$
where $X_1$ is the first covariate in $\bm X$ and $\bm X_{-1}$ is the covariate vector of $\bm X$ excluding $X_1$. Consider the following estimating equation for $\beta_1^*$,
\begin{eqnarray}\label{eq:estimating_equation_beta_1}
	E\left[\left\{S(\bm \beta;\pi, \widehat{Q})\right\}^\top \bm v\right],
\end{eqnarray}
where $\bm v^\top=(1, -{\bm w^*}^\top)$. By using this estimating equation with $\bm\beta=(\beta_1,\widehat{\bm\beta}_{-1})$, the estimation error of $\widehat{\bm\beta}_{-1}$ will not affect the estimation of $\beta_1^*$.

In order to remove the estimation error of $\widehat{Q}$, we will choose a specific weighting function. Under the proposed estimating equation~\eqref{eq:estimating_equation_beta_1}, for any choice of $\pi$, the first-order bias of the proposed estimating equation~\eqref{eq:estimating_equation_beta_1} with $Q$ being replaced by $\widehat{Q}$ is 
\begin{eqnarray*}
	&&E\left[\left\{R/\pi(\widetilde{\bm X})-1\right\}\left\{\widehat{g}(\widehat{\bm \Gamma}^\top \widetilde{\bm X}) -g(\bm \Gamma^\top \widetilde{\bm X})\right\}\bm X^\top \bm v\right]\\
 &\approx&E\left[\left\{R/\pi(\widetilde{\bm X})-1\right\}\left\{\widehat{g}(\bm \Gamma^\top \widetilde{\bm X}) -g(\bm \Gamma^\top \widetilde{\bm X})\right\}\bm X^\top \bm v\right].
\end{eqnarray*}
 In order to remove the estimation error of $\widehat{Q}$, one possible strategy is to choose $\pi(\widetilde{\bm X})$ such that
\begin{eqnarray}\label{eq:required_equation}
 E\left[\left\{R/\pi(\widetilde{\bm X})-1\right\}f(\bm \Gamma^\top \widetilde{\bm X})\bm X^\top\bm v\right]=0,\text{ for any } f\in L_2(\bm \Gamma^\top \widetilde{\bm X}).
\end{eqnarray}

Let $P(R=1)=\rho>0$ and $\eta(\cdot\mid R)$ be the conditional density function of ${\bm \Gamma}^\top \widetilde{\bm X}$ given $R$. Define $$J_r({\bm \Gamma}^\top \widetilde{\bm x})=E\left[\bm X^\top v\mid {\bm \Gamma}^\top \widetilde{\bm X}={\bm \Gamma}^\top \widetilde{\bm x}, R=r\right]\eta({\bm \Gamma}^\top \widetilde{\bm x}\mid R=r),$$ for $r=0$ and $1$. Theorem~\ref{thm:solution_to_equation} characterizes the solution to Equation~\eqref{eq:required_equation}. The proof of Theorem~\ref{thm:solution_to_equation} can be found in the Online Supporting Information.
\begin{theorem}\label{thm:solution_to_equation}
We assume the following regularity condition: for almost all $\widetilde{\bm x}$, if $J_1({\bm \Gamma}^\top \widetilde{\bm x})=0$, then $J_0({\bm \Gamma}^\top \widetilde{\bm x})=0$. Then 
$$
    \pi_*^{-1}(\widetilde{\bm x})=E\left[\bm X^\top\bm v\mid {\bm \Gamma}^\top \widetilde{\bm X}\right]/E\left[R\bm X^\top\bm v\mid {\bm \Gamma}^\top \widetilde{\bm X}\right]
$$
is well-defined. Further, the solution set of Equation~\eqref{eq:required_equation} can be characterized as all the functions of the form $\pi^{-1}_*(\widetilde{\bm x})+\mathcal{T}h(\widetilde{\bm x})$ on event $\{\widetilde{\bm x}:J_1({\bm \Gamma}^\top \widetilde{\bm x})\not=0\}$, where $\mathcal{T}$ is a linear operator defined as
    $$
        \mathcal{T}h=h-E\left[h \bm X^\top \bm v\mid {\bm \Gamma}^\top \widetilde{\bm X}, R=1\right]/E\left[\bm X^\top v\mid {\bm \Gamma}^\top \widetilde{\bm X}, R=1\right]
    $$
    and $h$ is an arbitrary function in $L_2(\widetilde{\bm X})$.
\end{theorem}

\begin{remark}
    The term $\pi_*^{-1}(\widetilde{\bm x})$ can also be written as $$1+\left\{\rho J_1({\bm \Gamma}^\top \widetilde{\bm x})\right\}^{-1}J_0({\bm \Gamma}^\top \widetilde{\bm x})(1-\rho).$$ This formulation indicates that we can estimate $J_1$, $J_0$, and $\rho$, and then use these estimates and $\widehat{\bm\Gamma}$ to construct an estimator for $\pi_*^{-1}(\widetilde{\bm x})$.
\end{remark}

\begin{remark}

The regularity condition is required to ensure that $\pi^{-1}_*$, as well as the characterization in Theorem~\ref{thm:solution_to_equation}, are both well-defined. This condition can be easily satisfied. For example, assume $\eta({\bm \Gamma}^\top \widetilde{\bm X}\mid R=1)>0$. If $$E\left[\bm X^\top v\mid {\bm \Gamma}^\top \widetilde{\bm X}, R=1\right]$$ has a continuous distribution, then we have 
$
P\left[J_1({\bm \Gamma}^\top \widetilde{\bm X})=0\right]=0,
$
and the regularity condition is naturally satisfied. In addition, when $P(R=1\mid Z, \bm X)=P(R=1\mid {\bm \Gamma}^\top \widetilde{\bm X})$, the regularity condition is also satisfied because
$$
    E\left[\bm X^\top \bm v\mid {\bm \Gamma}^\top \widetilde{\bm X}, R=1\right]=E\left[\bm X^\top \bm v\mid {\bm \Gamma}^\top \widetilde{\bm X}, R=0\right].
$$
\end{remark}

\begin{remark}
The term $\pi_*^{-1}(\widetilde{\bm x})$ is a function of $\bm \Gamma^\top\widetilde{\bm x}$. This can be interpreted as the consequence of $Y\perp R\mid \bm \Gamma^\top\widetilde{\bm X}$. However, the term $\pi^{-1}_*(\widetilde{\bm x})+\mathcal{T}h(\widetilde{\bm x})$ may not include $P(R=1\mid \bm \Gamma^\top\widetilde{\bm X})$. This implies that the true propensity based on $\bm \Gamma^\top\widetilde{\bm X}$ may not be sufficient to remove the bias.
\end{remark}

We then use $\pi_*^{-1}(\widetilde{\bm X})$ to replace the $\pi^{-1}(\widetilde{\bm X})$ in estimating equation~\eqref{eq:estimating_equation_beta_1}. The weighting function $\pi_*^{-1}(\widetilde{\bm X})$ only depends on $\bm \Gamma^\top\widetilde{\bm X}$ and thus is a low-dimensional function. To estimate the weighting function $\pi_*^{-1}(\widetilde{\bm X})$, we consider the trimmed kernel estimates. First, we estimate $\bm w$ by minimizing
\begin{eqnarray*}
	\widehat{E}_n\left[b^{''}(\bm X^\top\widehat{\bm \beta})(X_1-X_{-1}^\top \bm w)^2\right],
\end{eqnarray*}
and construct $\widehat{\bm v}^\top=(1,-\widehat{\bm w}^\top)$. Then, using kernel regressions, we consider the following estimator for $\pi_*^{-1}$:
\begin{eqnarray*}
	\widehat{\pi}^{-1}(\widetilde{\bm x}; \widehat{\bm \Gamma}, \widehat{\bm v})=\begin{cases}
		1+\left\{\widehat{J}_1(\widehat{\bm \Gamma}^\top \widetilde{\bm x})\widehat{\rho}\right\}^{-1}\widehat{J}_0(\widehat{\bm \Gamma}^\top \widetilde{\bm x})(1-\widehat{\rho}) & \left|\widehat{J}_1(\widehat{\bm \Gamma}^\top \widetilde{\bm x})\right|> c_n,\\
		\widehat{\rho}^{-1}, & \left|\widehat{J}_1(\widehat{\bm \Gamma}^\top \widetilde{\bm x})\right|\leq c_n,
		\end{cases}
\end{eqnarray*}
where 
$
	\widehat{J}_r(\widehat{\bm \Gamma}^\top \widetilde{\bm x})=\widehat{E}_n\left[\bm X^\top \widehat{\bm v}K_{\hbar}(\widehat{\bm \Gamma}^\top \widetilde{\bm X}-\widehat{\bm \Gamma}^\top \widetilde{\bm x})\mid R=r\right]$, $\widehat{\rho}=\widehat{E}_n[R]$,$
	K_{\hbar}(\cdot)=K(\cdot)/\hbar^{d},
$
and $\widehat{E}_n\left[\cdot\mid R=r\right]$ is the empirical mean over the samples with $R=r$.
The function $K(\cdot)$ is a kernel function with the order of $\nu$, and the bandwidth parameter $\hbar$ is selected according to Theorem~\ref{thm:asymptotic_normal}. The proposed estimator equals to the kernel regression when $\left|\widehat{J}_1(\widehat{\bm \Gamma}^\top \widetilde{\bm x})\right|$ is far from $0$, and equals to $\widehat{\rho}^{-1}$, when $\left|\widehat{J}_1(\widehat{\bm \Gamma}^\top \widetilde{\bm x})\right|$ is close to $0$. The term $c_n$ is used to trim possible extremities of the kernel regression estimates.

After obtaining the estimator $\widehat{\pi}^{-1}$ for $\pi_*^{-1}(\widetilde{\bm X})$, we construct the estimating equation by incorporating $\widehat{\pi}^{-1}$, i.e.,
$
    \widehat{E}_n\left[\left\{S({\bm \beta}; \widehat{Q}, \widehat{\pi})\right\}^\top\widehat{\bm v}\right]
$
with a constraint $\bm \beta_{-1}=\widehat{\bm \beta}_{-1}$. To avoid possible computational issues if directly solving this estimating equation (Chapter 5 in \citet{van2000asymptotic}), we use its first-order expansion and construct a one-step debiased estimator 
$
    \widetilde{\beta}_1=\widehat{\beta}_1-\widebar{I}^{-1}S,
$
where 
$
    S=\widehat{E}_n\left[\left\{S(\widehat{\bm \beta}; \widehat{Q}, \widehat{\pi})\right\}^\top\widehat{\bm v}\right],\quad $
    and $
			\widebar{I}=\widehat{E}_n\left[b^{''}(\bm X^\top\widehat{\bm \beta})X_1 \bm X^\top \widehat{\bm v}\right].
$
Another challenge in constructing the debiased estimator is that, the estimation errors of $\widehat{Q}$, $\widehat{\pi}$, and the samples used to construct the estimator are correlated. We adopt the cross-fitting procedure proposed in \cite{victor2018} in the implementation. 

\subsection{Implementation}
\label{sec:algorithm}

The entire procedure can be separated into two steps. In the first step, using all fully observed data, we obtain $\widehat{\bm \Gamma}$ and then,  regress $Y$ on $\widehat{\bm \Gamma}^\top\widetilde{\bm X}$ using kernel regressions and denote the estimated link function as $\widehat{g}$.  Using the estimated imputation model, $\widehat{Q}(Z, \bm X)=\widehat{g}(\widehat{\bm \Gamma}^\top\widetilde{\bm X})$,  we obtain an initial estimate $\widehat{\bm \beta}$. Using the initial estimate, we solve
$$
	\min_{\bm w} \widehat{E}_n\left[b^{''}(\bm X^\top\widehat{\bm \beta})(X_1-X_{-1}^\top \bm w)^2\right],
$$
and denote its minimizer as $\widehat{\bm w}$. Then we can construct $\widehat{\bm v}^\top=(1,-\widehat{\bm w}^\top)$. In the second step, we estimate the identified weighting function and use it to form a one-step debiased estimator for $\beta_1^*$. First, we split the entire data into $K$ subsets ($I_1, \ldots, I_K$) with equal sample sizes. For a specific set $k$, the estimated link function denoted as $\widehat{g}_{(-k)}(\cdot)$ is obtained through kernel regression of $Y$ w.r.t $\widehat{\Gamma}\widetilde{\bm X}$ using the data excluding $I_k$. The estimated weighting function denoted as $\widehat{\pi}_{(-k)}^{-1}(\widetilde{\bm X}; \widehat{\bm \Gamma}, \widehat{\bm v})$ is obtained through the truncated kernel regression using the data excluding $I_k$. Specifically,
\begin{eqnarray*}
	\widehat{\pi}_{(-k)}^{-1}(\widetilde{\bm x}; \widehat{\bm \Gamma}, \widehat{\bm v})=\begin{cases}
		1+\left\{\widehat{J}_1^{(-k)}(\widehat{\bm \Gamma}^\top \widetilde{\bm x})\widehat{\rho}\right\}^{-1}\widehat{J}_0^{(-k)}(\widehat{\bm \Gamma}^\top \widetilde{\bm x})(1-\widehat{\rho}) & \left|\widehat{J}_1^{(-k)}(\widehat{\bm \Gamma}^\top \widetilde{\bm x})\right|> c_n,\\
		\widehat{\rho}^{-1}, & \left|\widehat{J}_1^{(-k)}(\widehat{\bm \Gamma}^\top \widetilde{\bm x})\right|\leq c_n,
		\end{cases}
\end{eqnarray*}
where $
	\widehat{J}_r^{(-k)}(\widehat{\bm \Gamma}^\top \widetilde{\bm x})=\widehat{E}_n^{(-k)}\left[\bm X^\top \widehat{\bm v}K_{\hbar}(\widehat{\bm \Gamma}^\top \widetilde{\bm X}-\widehat{\bm \Gamma}^\top \widetilde{\bm x})\mid R=r\right]$ and $\widehat{E}_n^{(-k)}[\cdot\mid R=r]$ is the empirical average over the samples with $R=r$ and excluding those in $I_k$. Then, the one-step debiased estimator is
$
	\widetilde{\beta}_1=\widehat{\beta}_1-\widebar{I}^{-1}\widebar{S},
$
where 
$$
			\widebar{S} = \sum_{k=1}^K {S}^{(k)}/K,\quad {S}^{(k)} = \widehat{E}_n^{(k)}\left[\left\{S(\widehat{\bm \beta}; \widehat{Q}_{(-k)}, \widehat{\pi}_{(-k)})\right\}^\top\widehat{\bm v}\right].
$$
A summary of the entire algorithm can be found in the Online Supporting Information. To estimate the asymptotic variance of $\widetilde{\beta}_1$, we bootstrap based on the entire sample for $B$ times; for the $b$th bootstrapped dataset, we implement the algorithm and obtain $\widetilde{\beta}_1^{(b)}$, where $b=1,\cdots, B$. We use the variance of $\left\{\widetilde{\beta}_1^{(b)}\right\}_{b=1}^B$ as the estimate for the asymptotic variance to construct interval estimations.

\section{Theoretical properties}
\label{sec:theory}

In this section, we provide the asymptotic property of the proposed estimator. To accommodate the situation where the marginal missing rate may be close to $1$, we assume that the distribution of $(\bm X, Z)$ and the conditional distribution $Y\mid Z, \bm X$ do not depend on $n$; the missing propensity $P(R=1\mid Z,\bm X)$ may depend on $n$. Specifically, we consider two scenarios: 1) the missing propensity $P(R=1\mid Z,\bm X)$ does not change with $n$; 2) $P(R=1\mid Z,\bm X)=\rho_n w(\widetilde{\bm X})$ with $\rho_n\to 0$, where $w(\widetilde{\bm X})$ does not depend on $n$, $w(\widetilde{\bm X})$ is always bounded away from $\infty$, and $E\left[w(\widetilde{\bm X})\right]=1$. Notice that for both scenarios, we will only assume the relaxed positivity assumption: $P(R=1\mid \bm\Gamma^\top\widetilde{\bm X})>0$. This is a benefit of not using the inverse of the true propensity $P(R=1\mid Z, \bm X)$ in the estimation. For simplicity, we focus on the required assumptions and theoretical results for Scenario 1) in the main text, and leave those for Scenario 2) in the Online Supporting Information.
In addition, the proofs of the theorems can be found in the Online Supporting Information. For Scenario 1), the following assumptions are required.
\begin{assumption} 
\label{cond:1}
The covariate $\bm X$'s and the surrogate outcome $Z$ are bounded, and the function $b^{''}(\cdot)$ is continuously differentiable; $\max\{\|\bm \beta^*\|_2,\|\bm v\|_2\}$ is bounded.
\end{assumption}
\begin{assumption} 
\label{cond:2}
There is a positive constant $\gamma_d>1/4$ such that
$$
	\|\widehat{g}(\widehat{\bm \Gamma}^\top\widetilde{\bm X})-Q(Z,\bm X)\|_{\infty}=O_p(n^{-\gamma_d}),
$$
and
$$
	\left\{\mathrm{vec}(\widehat{\bm \Gamma})-\mathrm{vec}({\bm \Gamma})\right\}=n^{-1}\sum_{i=1}^n 1\{R_{i}=1\}\psi(\bm X_{i}, Z_{i}, Y_{i})+o_p(n^{-1/2}),
$$
where $\psi(\bm X_{i}, Z_{i}, Y_{i})$ is bounded and $\mathrm{vec}(\cdot)$ represents the vectorization of the matrix. In addition, we assume that
$$
\sup_{\widetilde{\bm x}}\left|\left(\widehat{g}-g\right)(\widehat{\bm \Gamma}^\top\widetilde{\bm x})-\left(\widehat{g}-g\right)({\bm \Gamma}^\top\widetilde{\bm x})\right|=o_p(n^{-1/2}).
$$
\end{assumption}
\begin{assumption}
\label{cond:3} 
Function $J_r(\bm \Gamma^\top \widetilde{\bm x})$'s are $\nu$th order differentiable w.r.t $\bm \Gamma^\top \widetilde{\bm x}$ with bounded derivatives. Define $G(\bm \Gamma^\top \widetilde{\bm x})=J_1^{-1}(\bm \Gamma^\top \widetilde{\bm x})J_0(\bm \Gamma^\top \widetilde{\bm x})$. We assume that $G(\bm \Gamma^\top \widetilde{\bm x})$ is bounded away from $+\infty$ on the open set $\left\{\widetilde{\bm x}:J_1(\bm \Gamma^\top \widetilde{\bm x})\not = 0\right\}$. We also assume that the density function of $\bm A^\top\widetilde{\bm X}$, $\eta(\bm A^\top\widetilde{\bm X})$, is bounded away from $0$ and $+\infty$, and $\nu$th order differentiable with bounded derivatives.
\end{assumption}
\begin{assumption}
\label{cond:4} 
When $t>0$ is small enough, there exist positive constants $A_0$ and $\gamma_m$ such that
$
	P\left\{0\not= \left|E[\bm X^\top\bm v\mid \bm\Gamma^\top\widetilde{\bm X},R=1]\right|\leq t\right\}\leq A_0 t^{\gamma_m}.
$
\end{assumption}
\begin{assumption}
\label{cond:5} 
Take $c_n=\widetilde{\delta}_n^{2/(2+\gamma_m)}$, where $\widetilde{\delta}_n=(n\hbar^{d}/ \log n)^{-1/2}+\hbar^{\nu}+n^{-\gamma_d}$. We assume that $
    n^{-\gamma_d}\widetilde{\delta}_n^{\gamma_m/(2+\gamma_m)}=o(n^{-1/2}).
$
\end{assumption}

In Assumption~\ref{cond:1}, for ease of exposition, we assume a bounded design for each $\bm X$ and $Z$. Assumption~\ref{cond:2} includes the requirement for the chosen dimension reduction method to estimate $\bm \Gamma$ and the chosen  method to estimate $g$. Specifically, we assume that the uniform convergence rate of $\widehat{Q}$ is $O_p(n^{-\gamma_d})$, and the estimated subspace $\widehat{\bm \Gamma}$ is asymptotically linear. In addition, we assume that 
$
    \sup_{\widetilde{\bm x}}\left|\left(\widehat{g}-g\right)(\widehat{\bm \Gamma}^\top\widetilde{\bm x})-\left(\widehat{g}-g\right)({\bm \Gamma}^\top\widetilde{\bm x})\right|=o_p(n^{-1/2}).
$
Many dimension reduction methods and nonparametric methods satisfy Assumption~\ref{cond:2}. An example is given in the Online Supporting Information. Assumption~\ref{cond:3} assumes the regularity conditions to ensure that $\pi_*^{-1}$ can be well-estimated and the asymptotic variances are well-defined. Assumption~\ref{cond:4} restricts the concentration near $E[\bm X^\top\bm v\mid \bm\Gamma^\top\widetilde{\bm X},R=1]=0$ by the parameter $\gamma_m$. When $E[\bm X^\top\bm v\mid \bm\Gamma^\top\widetilde{\bm X},R=1]$ is bounded away from $0$, we have $\gamma_m=+\infty$; when $E[\bm X^\top\bm v\mid \bm\Gamma^\top\widetilde{\bm X},R=1]$ has a continuous distribution, we have $\gamma_m\geq 1$. Assumption~\ref{cond:5} specifies the condition on $\gamma_d$ and $\gamma_m$, which requires that $\gamma_d >(2+\gamma_m)/4(1+\gamma_m)$.

Under these assumptions, Theorem~\ref{thm:asymptotic_normal} shows that the one-step debiased estimator is asymptotically normal.
\begin{theorem}\label{thm:asymptotic_normal}
	Under Assumptions~\ref{cond:1}-~\ref{cond:5}, we have 
	$
		\sqrt{n}(\widetilde{\beta}_1-\beta_1^*)\to N(0,\sigma^2),
	$
	where the formula of the asymptotic variance $\sigma^2$ can be found in the Online Supporting Information.
\end{theorem}

Theorem~\ref{thm:asymptotic_normal} implies that we can construct a valid confidence interval if we can consistently estimate $\sigma^2$. To estimate the asymptotic variance $\sigma^2$, we can adopt a plug-in approach. When the asymptotic variance of $\widehat{\bm \Gamma}$ is explicitly known, we can construct estimators for the unknown parameters in the asymptotic variance formula and construct a plug-in estimator for $\sigma^2$. In this work, we choose the bootstrap procedure, which has been shown to have a better numerical performance for regression with semi-nonparametric nuisance models \citep{liu2020semi}.

In addition, Corollary~\ref{corollary:efficient} shows a sufficient condition that the semiparametric lower bound can be achieved.
\begin{corollary}\label{corollary:efficient}
When the true propensity $P(R=1\mid Z, \bm X)=P(R=1\mid \bm \Gamma^\top\widetilde{\bm X})$, the one-step debiased estimator $\widetilde{\beta}_1$ obtains the semiparametric lower bound.
\end{corollary}

\section{Simulations}
\label{sec:sim}
 In this section, we conduct simulations and compare the proposed method with other methods to demonstrate 1)  the advantage of avoiding complex propensity estimation; and 2) the efficiency gain from incorporating the surrogate outcome. To show the advantage of avoiding modeling the complex propensity, we compare our proposed method with two baseline approaches. Baseline 1 follows the double machine learning procedure proposed in \cite{victor2018}, which estimates both the propensity and the imputation model using kernel regressions. When using the kernel regressions to estimate the propensity and the imputation model, we first implement dimension reduction and then conduct the kernel regression. Another baseline approach (Baseline 2) adopts the same procedure as Baseline 1 but uses a logistic regression to estimate the missing propensity. For both Baselines 1 and 2, we implement a threshold of $0.001$ for the estimated propensities to avoid extreme value. To show the efficiency gain from incorporating the surrogate outcome, besides the proposed procedure using the surrogate outcome (denoted as ``Proposed with $Z$''), we implement another approach (denoted as ``Proposed w/o $Z$'') following the same procedure but only using  $\bm X$ (no surrogate outcome $Z$) in the dimension reduction, imputation model estimation, and weighting function estimation. For both proposed procedures, we specify $c_n$ following the results in Theorem~\ref{thm:asymptotic_normal}. For the dimension reduction adopted in all these approaches, we use the kernel sliced regression method and choose the reduced dimension using cross-validation.

To compare with the proposed method, we consider $8$ simulation scenarios in total with varying missing rates, varying sample sizes, and different types of outcomes, i.e., continuous outcomes and binary outcomes. For each type of outcome, we consider a moderate marginal missing rate of 50\% and a high marginal missing rate of 90\%. For both scenarios, we change the sample size from $500$ to $1000$. To generate the data under each scenario, we first generate the missing indicator $R$ following a Bernoulli distribution with the success probabilities of $0.5$ (moderate marginal missing rate) or $0.1$ (high marginal missing rate). Then, we generate the covariate based on $R=1$ or $R=0$. When the outcome is missing ($R=0$), the covariates, i.e., $\bm X\mid R =0$, follow a standard multivariate Gaussian distribution with zero means and the identity covariance matrix; when the outcome is observed ($R=1$), the covariates follow a mixture of two multivariate Gaussian distributions. With a probability of $0.7$, the covariates are generated following a standard multivariate Gaussian distribution; otherwise, the covariates are generated following a multivariate Gaussian distribution $N(1,1.5 \bm I_p)$, i.e., $\bm X\mid R=1 \sim \xi N(0, I_p) + (1-\xi) N(1, 1.5 \bm I_p)$, where $\xi$ follows a Bernoulli distribution with a success probability of $0.7$. The surrogate outcome is generated from $Z = \delta \bm X^\top \bm \beta_0 + \sum_{j=5}^8 |X_j|/4 + \epsilon_z$ where $\epsilon_z \sim N(0,1)$. To generate the outcome $Y$ given the covariates, for the scenario with continuous outcomes, we consider
$
Y = \bm X^\top \bm \beta_0/2 + Z + \epsilon,
$
where $\bm \beta_0 = (1,1,-1,-1,0,0,0,0)^\top$ and $\epsilon \sim N(0,1)$. 
For the scenario with binary outcomes, we consider
$
Y = 1\{\bm X^\top \bm \beta_0/4 + Z + \epsilon>0\}.
$
 For the scenario with continuous outcomes, $\delta$ is fixed at $0.5$; for the scenario with binary outcomes, $\delta$ is fixed at $0.25$. 

To evaluate the proposed methods, we compare the coverage and deviance in all the scenarios. For each scenario, we run $500$ replicates. For each replicate, we estimate the coefficients and use bootstrapping ($500$ bootstraps) to construct a $95\%$-confidence interval using the training samples. We also calculate the deviance using the coefficients estimated by each approach on an independently generated testing dataset with a sample size of $10^4$.

\begin{figure}[!hbt]
\centering

\includegraphics[scale=0.45]{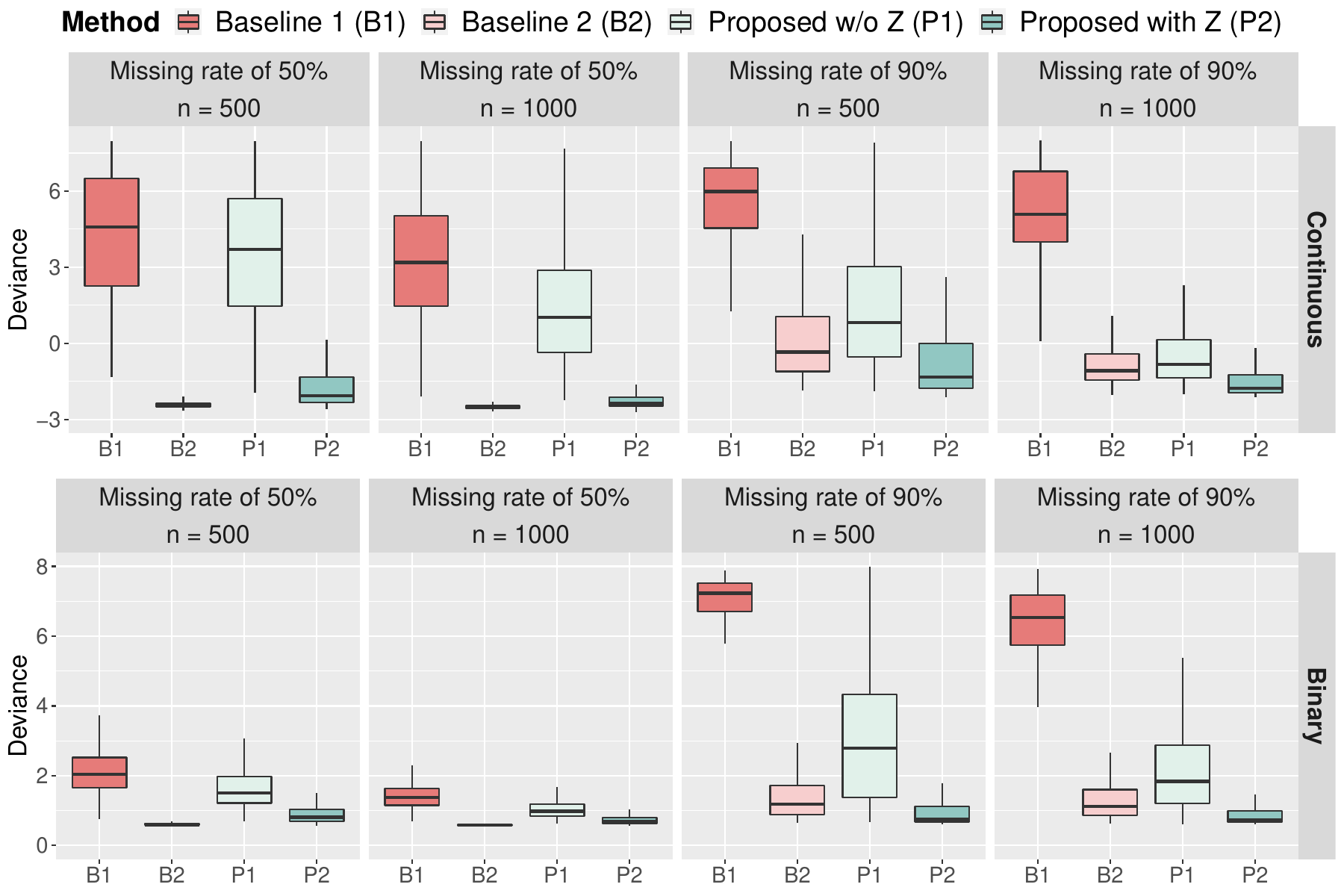}
\caption{Deviance under different missing rates, sample sizes, and outcome types.\label{fig:deviances}}
\end{figure}

Table~\ref{tbl:coverages} exhibits the coverage metrics of the first four coefficients. From Table~\ref{tbl:coverages}, the proposed methods (Proposed w/o and with $Z$) achieve the nominal coverage in all the scenarios, whereas Baseline 1 and Baseline 2 do not, especially in high missing rate settings. This suggests both Baseline 1 and Baseline 2 are incapable of handling the complex missing mechanism. Figure \ref{fig:deviances} summarizes the deviance metrics for all  different scenarios. At the moderate missing rate (the left two columns), the method Proposed with $Z$ performs comparably to Baseline 2 but outperforms the other methods. In the high missing rate scenarios, Proposed with $Z$ achieves the minimum deviance. When comparing the two  methods, Proposed with $Z$ surpasses Proposed w/o $Z$ in all the scenarios due to the efficiency gain from $Z$. In summary, considering both coverage and deviance metrics, the proposed method with the surrogate outcome is dominant.

\begin{table}
\caption{Coverage of the $95$\% confidence interval for the coefficients.}
\centering
\begin{tabular}{lllllllll}

                  & \multicolumn{4}{c}{$n=500$}                                                                                                   & \multicolumn{4}{c}{$n=1000$}                                                                                                  \\

                  & \multicolumn{8}{c}{Continuous, Missing rate of 50\%}                                                                                                                                                                                                          \\
                  & \multicolumn{1}{c}{$\beta_1$} & \multicolumn{1}{c}{$\beta_2$} & \multicolumn{1}{c}{$\beta_3$} & \multicolumn{1}{c}{$\beta_4$} & \multicolumn{1}{c}{$\beta_1$} & \multicolumn{1}{c}{$\beta_2$} & \multicolumn{1}{c}{$\beta_3$} & \multicolumn{1}{c}{$\beta_4$} \\
Baseline 1        & 0.872                         & 0.840                         & 0.886                         & 0.864                         & 0.847                         & 0.769                         & 0.861                         & 0.817                         \\
Baseline 2        & 0.896                         & 0.912                         & 0.906                         & 0.928                         & 0.885                         & 0.873                         & 0.893                         & 0.875                         \\
Proposed w/o $Z$  & 0.920                         & 0.938                         & 0.910                         & 0.922                         & 0.901                         & 0.901                         & 0.891                         & 0.893                         \\
Proposed with $Z$ & 0.952                         & 0.964                         & 0.958                         & 0.968                         & 0.954                         & 0.954                         & 0.956                         & 0.954                         \\
                  & \multicolumn{8}{c}{Continuous, Missing rate of 90\%}                                                                                                                                                                                                          \\
                  & \multicolumn{1}{c}{$\beta_1$} & \multicolumn{1}{c}{$\beta_2$} & \multicolumn{1}{c}{$\beta_3$} & \multicolumn{1}{c}{$\beta_4$} & \multicolumn{1}{c}{$\beta_1$} & \multicolumn{1}{c}{$\beta_2$} & \multicolumn{1}{c}{$\beta_3$} & \multicolumn{1}{c}{$\beta_4$} \\
Baseline 1        & 0.866                         & 0.840                         & 0.844                         & 0.854                         & 0.740                         & 0.798                         & 0.752                         & 0.762                         \\
Baseline 2        & 0.904                         & 0.876                         & 0.922                         & 0.920                         & 0.910                         & 0.898                         & 0.894                         & 0.902                         \\
Proposed w/o $Z$  & 0.954                         & 0.966                         & 0.968                         & 0.946                         & 0.942                         & 0.944                         & 0.938                         & 0.952                         \\
Proposed with $Z$ & 0.946                         & 0.968                         & 0.966                         & 0.962                         & 0.946                         & 0.964                         & 0.954                         & 0.956                         \\
                  & \multicolumn{8}{c}{Binary, Missing rate of 50\%}                                                                                                                                                                                                              \\
                  & \multicolumn{1}{c}{$\beta_1$} & \multicolumn{1}{c}{$\beta_2$} & \multicolumn{1}{c}{$\beta_3$} & \multicolumn{1}{c}{$\beta_4$} & \multicolumn{1}{c}{$\beta_1$} & \multicolumn{1}{c}{$\beta_2$} & \multicolumn{1}{c}{$\beta_3$} & \multicolumn{1}{c}{$\beta_4$} \\
Baseline 1         & 0.908                         & 0.918                         & 0.916                         & 0.932                         & 0.892                         & 0.882                         & 0.918                         & 0.900                         \\
Baseline 2         & 0.928                         & 0.932                         & 0.916                         & 0.930                         & 0.920                         & 0.898                         & 0.894                         & 0.912                         \\
Proposed w/o $Z$  & 0.968                         & 0.970                         & 0.956                         & 0.950                         & 0.962                         & 0.964                         & 0.964                         & 0.964                         \\
Proposed with $Z$ & 0.970                         & 0.972                         & 0.948                         & 0.962                         & 0.958                         & 0.968                         & 0.956                         & 0.946                         \\
                  & \multicolumn{8}{c}{Binary, Missing rate of 90\%}                                                                                                                                                                                                              \\
                  & \multicolumn{1}{c}{$\beta_1$} & \multicolumn{1}{c}{$\beta_2$} & \multicolumn{1}{c}{$\beta_3$} & \multicolumn{1}{c}{$\beta_4$} & \multicolumn{1}{c}{$\beta_1$} & \multicolumn{1}{c}{$\beta_2$} & \multicolumn{1}{c}{$\beta_3$} & \multicolumn{1}{c}{$\beta_4$} \\
Baseline 1         & 0.950                         & 0.940                         & 0.948                         & 0.962                         & 0.892                         & 0.942                         & 0.964                         & 0.940                         \\
Baseline 2         & 0.932                         & 0.930                         & 0.920                         & 0.956                         & 0.924                         & 0.912                         & 0.896                         & 0.926                         \\
Proposed w/o $Z$  & 0.956                         & 0.950                         & 0.948                         & 0.964                         & 0.954                         & 0.962                         & 0.944                         & 0.950                         \\
Proposed with $Z$ & 0.966                         & 0.966                         & 0.976                         & 0.948                         & 0.942                         & 0.962                         & 0.966                         & 0.974     \\
                   
\end{tabular}
\label{tbl:coverages}
\end{table}

\section{Application to PROMIS global physical health T-score}
\label{sec:real}
In this section, we applied our proposed method to predict whether the improvement of the PROMIS global physical health T-score will exceed the MCID after receiving total joint replacement using the information obtained before scheduling the surgery. In addition to making prediction, we also aimed to identify the driving factors of not achieving the MCID in the presence of the high missing-rate outcome. The dataset includes 1044 University of Florida Health patients, who participated in the pre-surgical survey and underwent the total joint replacement surgery. In the analysis, we incorporated many baseline covariates including demographics, socioeconomic characteristics, medical history, and care characteristics before surgery  (e.g., 30 days before admission for surgery). According to the convention of constructing MCID, the outcome of interest is obtained based on the one-half standard deviation of the difference between pre- and post-surgical PROMIS global health T-scores \citep{fontana2019can, katakam2022development}. 

For our data, the difference between pre- and post-surgical scores has an average of $9.5$ and a standard deviation of $8.1$, and consequently, the MCID is $4.1$. Thus, the outcome $Y = 1$ if the difference is less than $4.1$, and $Y = 0$, otherwise. In terms of the missing proportion, all the 1044 identified patients took the pre-surgical survey, but only 261 patients ($25\%$) responded to the post-surgical survey (the missing-rate is $75\%$). Of the patients who took both surveys, $67 (25.7\%)$ patients did not meet the MCID. Since the PROMIS global physical T-score is derived from the ten survey questionnaire items, individual items can be considered as candidates for the surrogate outcome. To construct an informative surrogate, we relied on actual data and regressed the target outcome w.r.t the pre-surgical survey responses, and used the predicted values as a single informative surrogate outcome.
    
We conducted two analyses to investigate the performance of the proposed method. In the first analysis, we compared the proposed methods with the baseline methods in terms of the deviance $E\left[\ell(\widehat{\bm \beta})\right]$, where $\ell(\widehat{\bm \beta})=b(\bm X^\top\widehat{\bm \beta})-\widehat{g}(\widehat{\bm \Gamma}^\top \widetilde{\bm X})\bm X^\top\widehat{\bm \beta}$. Specifically, we randomly split the entire dataset into a training dataset and a testing dataset with equal sample sizes. We estimated coefficients using the training dataset, and then we calculated the deviance on the testing dataset. This procedure was repeated 1000 times. In the second analysis, we fitted the model on the entire dataset and compare the variables selected by different methods.

Table~\ref{tab:compare_realdata} shows that the proposed method with $Z$ achieves the lowest deviance. In terms of the selected variables, compared with other methods, the proposed method with $Z$ uniquely revealed that geriatric patients were less likely to achieve the MCID (estimated coefficient is $0.367$; $95\%$-CI is $[0.020, 0.714]$). This is in accordance with the existing research finding that elderly patients were more likely to have post-operative adverse clinical outcomes than younger patients in total joint replacement \citep{higuera20112010, malkani2017high}. The coefficients with confidence intervals for other covariates are presented in the Online Supporting Information.

\begin{table}
\caption{Comparison of averaged deviances (standard deviations) in real data example}
\centering
	\begin{tabular}{cccc}
 
Baseline1            & Baseline2            & Proposed w/o $Z$       & Proposed w/ $Z$         \\
0.714 (0.105)        & 0.696 (0.094)        & 0.685 (0.091)        & 0.670 (0.094)        \\

\end{tabular}
\label{tab:compare_realdata}
\end{table}

\section{Discussion}
\label{sec:disc}

In this work, we propose a debias approach to estimating the parameters of interest under a possibly misspecified GLM. This approach uses an informative surrogate outcome which leads to a low-dimensional flexible imputation model, and estimates a low-dimensional weighting function instead of the complex propensity score. When the true propensity happens to enjoy the same low-dimensional structure, the proposed method achieves the semi-parametric efficiency lower bound. Compared with the double machine learning method, the proposed approach relaxes the requirement on the propensity estimation and maintains almost the same flexibility or requirement on the imputation model estimation. In addition, we relax the traditional positivity and only require an relaxed positivity assumption.

There are multiple future directions to extend the proposed approach. First, we can consider extending the proposed approach to high-dimensional settings where $p/n\to+\infty$. In a high-dimensional setting, the $\widehat{\bm \Gamma}$ may not be asymptotic normal due to the possible penalization. In this case, in order to achieve an asymptotic normal estimator, an additional debias procedure is needed to adjust for the bias due to the estimation error of $\widehat{\bm \Gamma}$. Second, we can investigate more choices of the function $h$. In this work, to pursue a low-dimensional weighting function, we choose $h$ such that $\mathcal{T}h=0$ when constructing the weighting function. However, we can choose other potential alternatives  to mitigate certain deficiencies such as a choice to minimize the asymptotic variance of the debiased estimator; a choice to avoid possible negative weights. Third, we can combine the proposed approach with the augmented minimax linear estimation \citep{hirshberg2021augmented} to avoid the computation of the Riesz representer. Especially in high-dimensional setting, an explicit form of the weights to debias $\widehat{\bm \Gamma}$ might be intractable.


\begin{center}
	{\large\bf Supplemental Materials}
\end{center}

\begin{description}
	
	\item[] \hspace{.65cm} Proofs of all theorems and additional simulation results are contained in the online supplemental materials. 
		
\end{description}

\bibliographystyle{apalike}
\bibliography{ref.bib}
\end{document}